%% file: singlehadkt_v1.5.tex
\RequirePackage{lineno} 
\documentclass[aps,prd,twocolumn,superscriptaddress,showpacs,preprintnumbers,amsmath,amssymb]{revtex4-1}
\usepackage{graphicx} 
\usepackage{dcolumn}  
\usepackage{rotating}
\usepackage{color}

\begin{document}



\title{ \quad\\[1.0cm]Transverse momentum dependent production cross sections of charged pions, kaons and protons produced in inclusive $e^+e^-$ annihilation at $\sqrt{s}=$ 10.58 GeV }
\include{pub531}
\noaffiliation
\begin{abstract}
We report measurements of the production cross sections of charged pions, kaons, and protons as a function of fractional energy, the event-shape variable called thrust, and the transverse momentum with respect to the thrust axis. These measurements access the transverse momenta created in the fragmentation process, which are of critical importance to the understanding of any transverse momentum dependent distribution and fragmentation functions. The low transverse momentum part of the cross sections can be well described by Gaussians in transverse momentum as is generally assumed but the fractional-energy dependence is non-trivial and different hadron types have varying Gaussian widths. The width of these Gaussians decreases with thrust and shows an initially rising, then decreasing fractional-energy dependence. The widths for pions and kaons are comparable within uncertainties, while those for protons are significantly narrower.    
These single-hadron cross sections and Gaussian widths are obtained from a $558\,{\rm fb}^{-1}$ data sample collected at the $\Upsilon(4S)$ resonance with the Belle detector at the KEKB asymmetric-energy $e^+ e^-$
collider.
\end{abstract}


\maketitle

\tighten

{\renewcommand{\thefootnote}{\fnsymbol{footnote}}}
\setcounter{footnote}{0}

Transverse-momentum-dependent parton distribution (PDF) and fragmentation functions (FF) have gained substantial interest due to the emergence of transverse spin dependent phenomena. The most well-known transverse-momentum-dependent effects in semi-inclusive deeply inelastic scattering (SIDIS) are related to the Sivers function \cite{sivers} and the Collins FF \cite{collins}. Both have been measured, first at HERMES \cite{hermes,hermes10}, and since confirmed by COMPASS \cite{compasscollins,Alekseev:2008aa,compass10} as well as by Belle, BaBar and BESIII in the case of the Collins FFs \cite{bellecollins,Seidl:2008xc,babarcollins,bes3}. They will play an important role in the future electron-ion collider to pin down the transverse momentum structure of the nucleon and its transverse spin (see \cite{eicwp} for details on the planned measurements). However, at present, for any transverse-momentum-dependent distribution (TMD) the explicit transverse momentum dependence is only poorly measured at best. The reason is that in most processes a convolution of several transverse momenta is involved (such as in SIDIS \cite{Airapetian:2012ki,Adolph:2013stb}), or that the statistical precision of several measurements is not sufficient, yet. Some direct access to the transverse momentum dependence of the polarized Collins FF's has been achieved by BaBar \cite{babarcollins,babark}, but little is known about the transverse momentum dependence of unpolarized fragmentation functions. Some old data for inclusive, unidentified hadron cross sections from $e^+e^-$ annihilation does exist \cite{Althoff:1983ew,Berger:1983yp}. An attempt at fitting these data was made, including the scale dependence \cite{elena}, but the precision of the data is very limited.
Also the SIDIS data has been fit by two groups \cite{Signori:2013mda,Angeles-Martinez:2015sea}, but the groups differ in their conclusions while explicit information for only the fragmentation part is necessary.
Furthermore, in order to relate the TMD effects in SIDIS and $e^+e^-$ to the large inclusive asymmetries measured in proton-proton collisions \cite{Abelev:2008af,Arsene:2008aa,Adare:2013ekj}, transverse momentum integrals over the TMDs are needed to arrive at the higher-twist functions relevant there \cite{Qiu:1998ia,Kanazawa:2000hz,Kanazawa:2014dca}.  
Another important aspect of TMDs that is yet to be addressed in more detail is the scale dependence, which is expected \cite{Bacchetta:2015ora} to be different from the collinear DGLAP \cite{Gribov:1972ri,Dokshitzer:1977sg,Altarelli:1977zs} evolution but again lacks data.\par
In this paper we present the unpolarized cross sections for single charged pion, kaon as well as proton production as a function of fractional energy, transverse momentum, and thrust where the reference axis is given by the thrust axis. These measurements are then related to the unpolarized single-hadron fragmentation functions $D_1^h(z,k_T,Q)$ with fractional energy $z=2E_{h}/\sqrt{s}$, and transverse momentum $k_T$ at the scale $Q = \sqrt{s}$.
Experimentally, the transverse momentum of the hadron is calculated relative to the thrust axis $\hat{\mathbf{n}}$ which maximizes the event-shape variable thrust $T$ \cite{Brandt:1964sa}:
\begin{equation}
T \stackrel{\mathrm{max}}{=} \frac{ \sum_h|\mathbf{P}^{\mathrm{CMS}}_h\cdot\mathbf{\hat{n}}|}{ \sum_h|\mathbf{P}^{\mathrm{CMS}}_h|}\quad.
\end{equation}
The sum extends over all detected particles, and $\mathbf{P}^{\mathrm{CMS}}_h$ denotes the momentum of particle $h$ in the center-of-mass system, CMS.\par
As the thrust variable describes how collimated all particles in an event are, the results are presented in bins of this value.

The paper is organized as follows: the detector setup and reconstruction criteria are detailed in Section I, in Section II the various corrections to get from the raw spectra to the final cross sections are discussed. In Section III the results are shown and compared to Monte-Carlo, MC, tunes before we proceed to study the transverse momentum behavior via Gaussian fits for small transverse momenta. We conclude with a summary in Section IV. (Note: additional figures and data files are available online in the supplement file \cite{supplement}.)
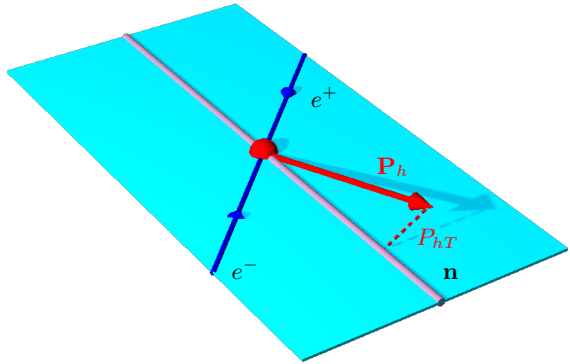
\begin{figure}[ht]
\begin{center}
  \includegraphics[width=8cm]{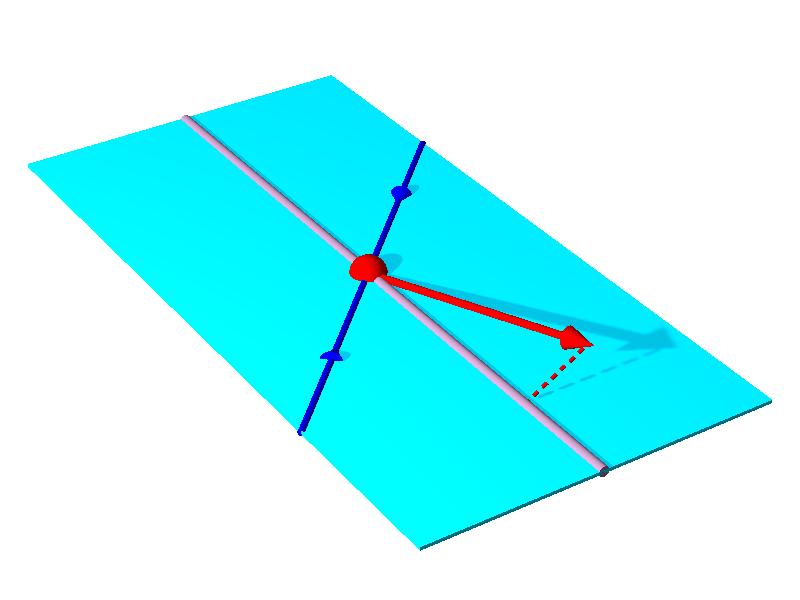}
  \put(-65,58){${\color{red}P_{hT}}$}
  \put(-80.,85.){${\color{red}\mathbf{P}_h}$}
  \put(-55.,45.){$\mathbf{n}$}
  \put(-105.,110.){$e^+$}
  \put(-135.,45.){$e^-$}
\caption{\label{fig:(h)x} Illustration of transverse-momentum-dependent single hadron fragmentation where the final-state hadron is depicted as a red arrow, the incoming leptons as blue arrows, and the event plane -- spanned by leptons (blue lines) and initial quarks/thrust axis $\mathbf{n}$ (purple line) -- is depicted as a light blue plane. The transverse momentum $P_{hT}$ is calculated relative to the thrust axis and depicted by the red, dashed line.}
\end{center}
\end{figure}

\section{Belle detector and data selection}

This single-hadron cross-section measurement is based on a data sample of $558\,{\rm fb}^{-1}$ 
collected with the Belle detector at the KEKB asymmetric-energy
$e^+e^-$ (3.5~GeV on 8~GeV) collider~\cite{KEKB,Abe:2013kxa}
operating at the $\Upsilon(4S)$ resonance (denoted as on-resonance), as well as a smaller data set taken 60 MeV below for comparison (denoted as continuum).

The Belle detector is a large-solid-angle magnetic
spectrometer that consists of a silicon vertex detector (SVD),
a 50-layer central drift chamber, an array of
aerogel threshold Cherenkov counters,  
a barrel-like arrangement of time-of-flight
scintillation counters, and an electromagnetic calorimeter
comprised of CsI(Tl) crystals located inside 
a superconducting solenoid coil that provides a 1.5~T
magnetic field.  An iron flux-return located outside of
the coil is instrumented to detect $K_L^0$ mesons and to identify
muons.  The detector
is described in detail elsewhere~\cite{Belle,Brodzicka:2012jm}.
A 1.5 cm beampipe with 1 mm thickness and a 4-layer
SVD and a small-cell inner drift chamber were used to record $558\,{\rm fb}^{-1}$ \cite{svd2}.  

The primary light ($uds$)- and charm-quark simulations used in this analysis were generated using {\sc pythia}6.2 \cite{pythia}, embedded into the EvtGen \cite{evtgen} framework, followed by a {\sc geant}3 \cite{geant} simulation of the detector response. The various MC samples were produced separately for light ($uds$) and charm quarks, and on the generator level several JETSET\cite{Sjostrand:1993yb} settings were produced in order to study their impact. For generator level MC to data comparisons, long-lived weak decays, which normally are handled in {\sc geant}, were allowed in EvtGen. In addition, we generated charged and neutral $B$ meson pairs from $\Upsilon(4S)$ decays in EvtGen, $\tau$ pair events with the KKMC \cite{Jadach:1999vf,Banerjee:2007is} generator and the {\sc Tauola} \cite{tauola} decay package, and other events with either {\sc pythia} or dedicated generators \cite{aafh} such as for two-photon processes. 
\subsection{Event and track selection}
The goal of this analysis is to extract hadron cross sections from $uds$ and charm pair events. Therefore events are required to have a visible energy of all detected charged tracks and neutral clusters above 7 GeV (to remove $\tau$ pair events) and either a heavy-jet mass (the greater of the invariant masses of all particles in a hemisphere as generated by the plane perpendicular to the thrust axis) above 1.8 GeV/c$^2$ or a ratio of the heavy-jet mass to visible energy above 0.25. Also, events need to have at least three reconstructed charged tracks, which reduces two-photon processes.
The thrust value is calculated as descibed above, where all detected particles and neutral clusters are included. For the charged particles, the mass hypothesis for the identified particle type is taken into account when boosting into the CMS. The thrust axis is required to point into the barrel part of the detector by having a $z$ component $|\hat{n}_z|<0.75$ in order to reduce the amount of thrust-axis smearing due to undetected particles in the forward/backward regions. 
Tracks are required to be within 4 cm (2 cm) of the interaction point along (perpendicular to) the positron beam axis. Each track is required to have at least three SVD hits and fall within the polar-angular acceptance of $-0.511  < \cos\theta_{\mathrm{lab}} < 0.842 $ in order to have PID information from all relevant PID detectors.
The fractional energy of each track is required to exceed 0.1 and the transverse momentum with respect to the thrust axis is then calculated in the CMS as illustrated in Fig.~\ref{fig:(h)x}. Also a minimum transverse momentum in the laboratory frame with respect to the beam axis of 100 MeV/c is imposed to ensure the particles traverse the magnetic field. 

\begin{figure*}[ht]
\begin{center}
\includegraphics[width=0.8\textwidth]{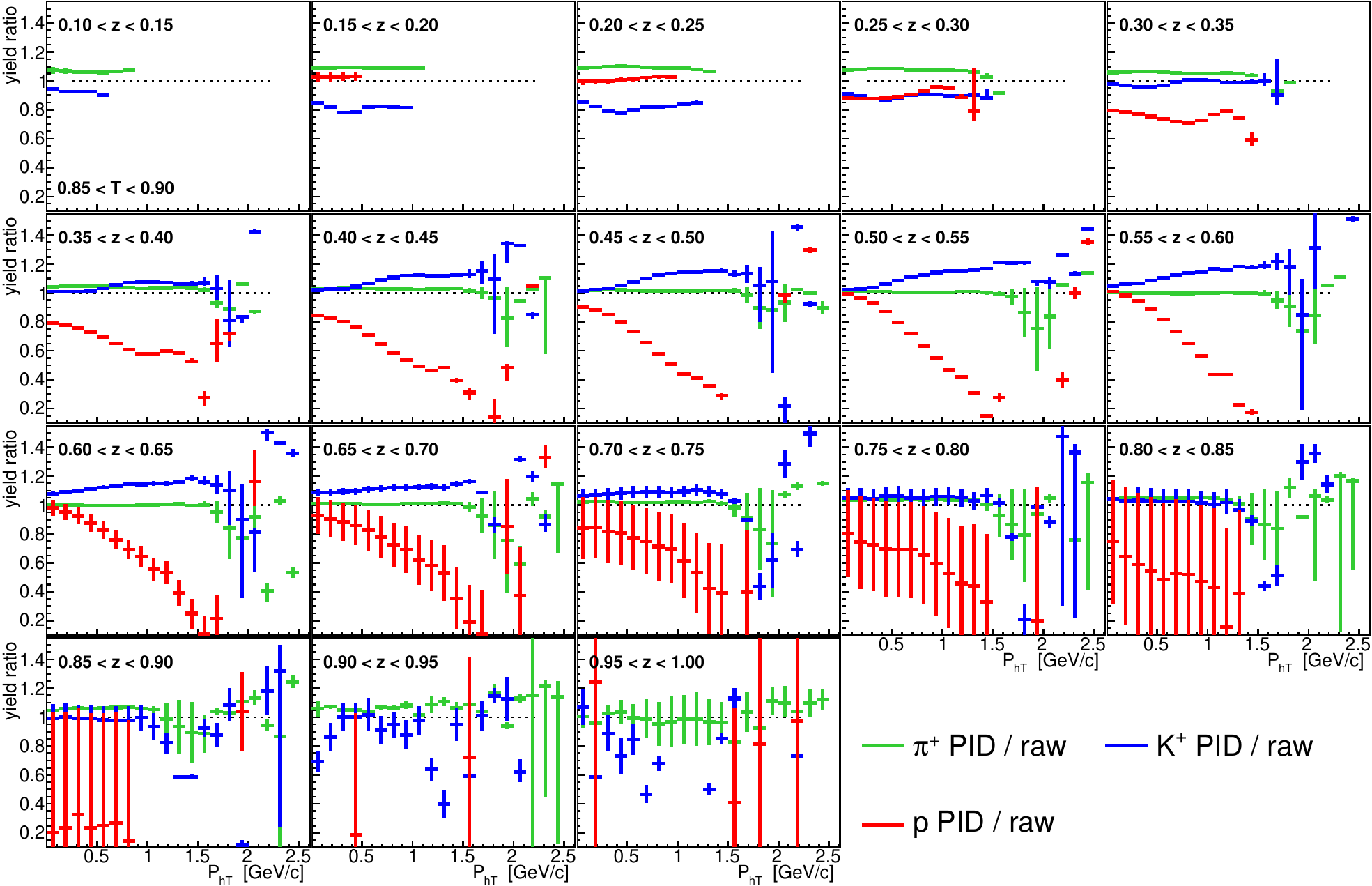} 
\caption{\label{fig:beforeafterpid} Ratio of yields after to before applying the PID correction for pions (green), kaons (blue), and protons (red), as a function of the transverse momentum $P_{hT}$ in bins of $z$ for an intermediate thrust bin. Empty bins are visible where the yields become zero, especially for high-$z$ bins as well as for kinematically inaccessible low-$z$ protons. The error bars represent the systematic uncertainties assigned for this correction step.}
\end{center}
\end{figure*}

\subsection{PID selection}
To apply the PID correction according to the PID efficiency matrices used in previous results \cite{martin}, the same selection criteria are applied first to define a charged track as a pion, kaon, proton, electron or muon. This information is determined from normalized likelihood ratios that are constructed from various detector responses.  
If the muon-hadron likelihood ratio is above 0.9, the track is identified as a muon. Otherwise, if the electron-hadron likelihood ratio is above 0.85, the track is identified as an electron. 
If neither of these applies, the track is identified as a kaon by a kaon-pion likelihood ratio above 0.6 and a kaon-proton likelihood ratio above 0.2. Pions are identified with the kaon-pion likelihood ratio below 0.6 and a pion-proton ratio above 0.2. Finally, protons are identified with kaon-proton and pion-proton ratios below 0.2. Here, and in the remaining sections of this presentation, protons will refer to combinations of protons and anti-protons unless the charge is explicitly mentioned. While neither muons nor electrons are considered explicitly for the single hadron analysis, they are retained as necessary contributors for the PID correction, wherein a certain fraction enters the pion, kaon, and proton samples under study.   

\section{Hadron analysis and corrections}
In the following sections, the hadron yields are extracted and, successively, the various corrections are applied and the corresponding systematic uncertainties are determined to arrive at the single hadron differential cross sections $d^3\sigma(e^+e^-\rightarrow hX)/dzdP_{hT}dT$ depending on fractional energy $z$, transverse momentum $P_{hT}$, and thrust value $T$. 
\subsection{Binning and cross-section extraction}
For the hadron cross section, a ($z$, $P_{hT}$) binning of 18 equidistant $z$ bins from 0.1 to 1.0 and 20 equidistant $P_{hT}$ bins from 0 to 2.5 GeV/c is chosen. The thrust values are separated into six bins with boundaries at $0.5,0.7,0.8,0.85,0.9,0.95$, and $1.0$. Due to the correlation between total hadron energy and transverse momentum, the range in $P_{hT}$ is kinematically limited at low $z$ bins. 

\begin{figure*}[htb]
\begin{center}
\includegraphics[width=0.9\textwidth]{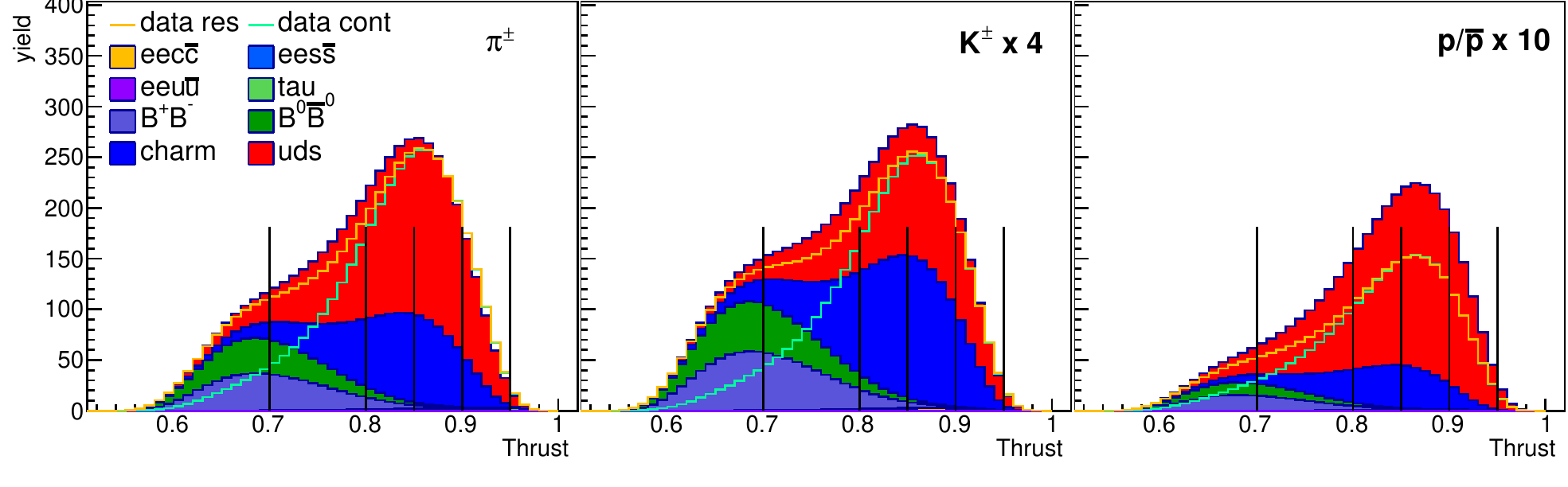}
\caption{\label{fig:nonqq:thrustdist} Contributions to the thrust distributions  from various processes for the reconstructed pion (left), kaon (center), and proton (right) yields at the $\Upsilon(4S)$ resonance. From bottom to top, the stacked contributions from $eec\bar{c}$ (yellow), $ees\bar{s}$ (dark blue), $eeu\bar{u}$ (purple), $\tau^+\tau^-$ (light green), $\Upsilon(4S)\rightarrow B^+B^-$ (violet), $\Upsilon(4S)\rightarrow B^0\bar{B}^0$ (dark green), charm (blue), and $uds$ (red) are shown. For comparison, the data for continuum (turquoise, denoted as ``data cont'') and on-resonance (orange, denoted as ``data res'') are also shown. The black vertical lines display the thrust bin boundaries used in this analysis.}
\end{center}
\end{figure*}

The distributions of thrust for the selected hadron samples are displayed in Fig.~\ref{fig:nonqq:thrustdist}, where the different processes are depicted. It can be seen that $uds$ and charm events peak at high thrust values, which is why in the following most corrections and results are displayed in the $0.85 <T<0.9$ thrust bin. The results of other bins are shown in the supplement file \cite{supplement}, as are logarithmic versions of the thrust contributions. 

\subsection{PID correction}

Following Ref.~\cite{martin}, particle misidentification is addressed in a very fine binning of $17$ laboratory momentum and $9$ polar angular bins. In each bin the particle misidentification matrix between true and detected particle types is reconstructed using five particle hypotheses (pions, kaons, protons, muons, and electrons) based on decays of $D^{*+}$, $\Lambda$, and $J/\psi$ from data where the actual particle type can be inferred from the decay chain. In the boundaries of the acceptance, MC information needs to be included to determine all matrix entries. These boundary bins are extrapolated either directly from the MC or by following the bins filled by data using only the behavior of the MC. The particle yields are then corrected using the inverse matrices and their uncertainties, and the uncertainties due to these MC extrapolations are assigned as systematic uncertainties.
The corrections have a moderate effect on the hadron yields, with slight increases of the pion yields and reductions of kaon yields at low $z$, mostly due to pion-kaon mis-identification. At higher $z$, kaon yields increase at the expense of proton yields with increasing transverse momentum. 
The ratios relative to the uncorrected hadron yields are shown in Fig.~\ref{fig:beforeafterpid} for an intermediate thrust bin. The behavior for other thrust bins is similar.
\begin{figure*}[htb]
\begin{center}
\includegraphics[width=0.9\textwidth]{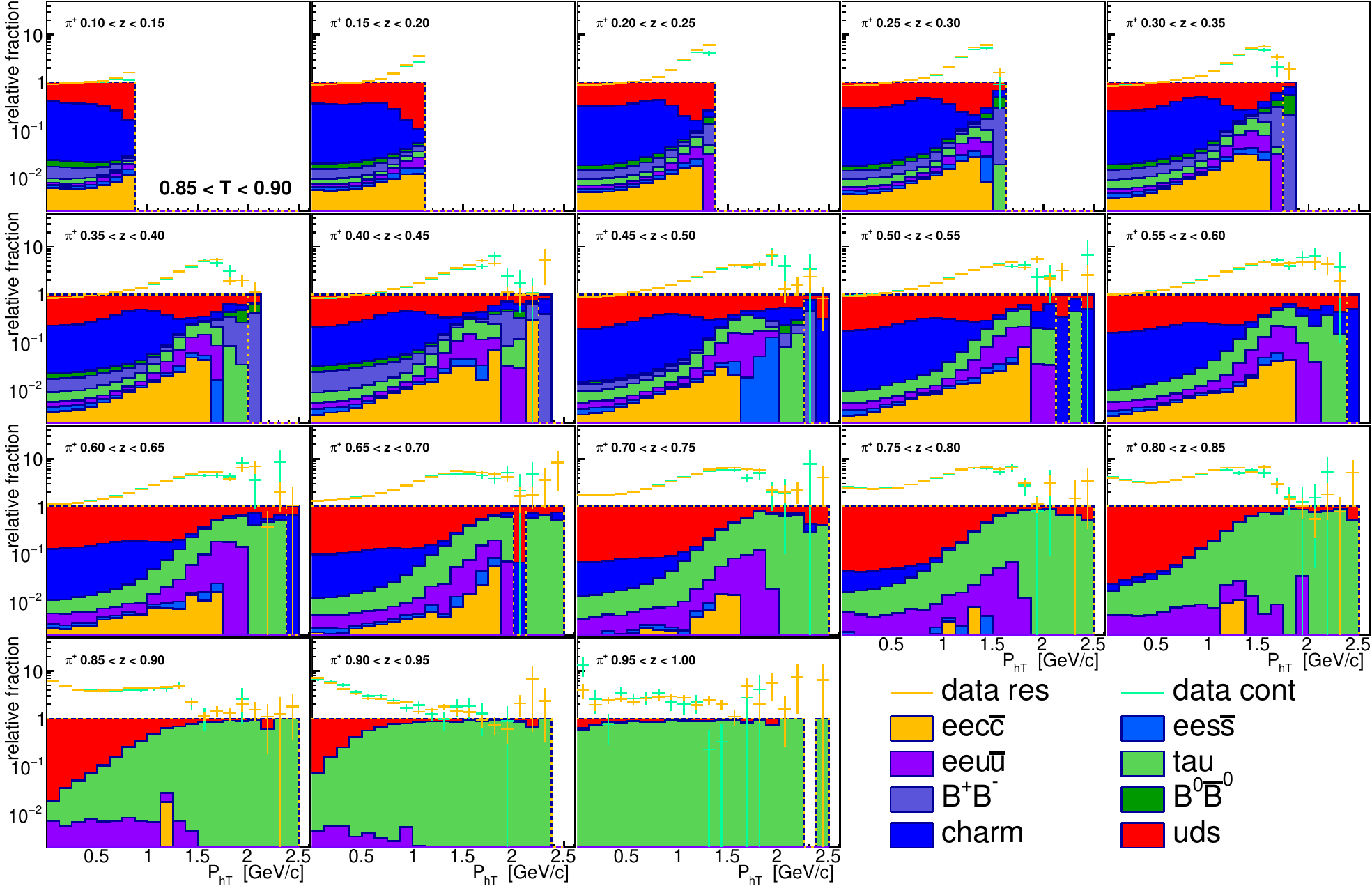}
\caption{\label{fig:nonqq:nonqqratio} Contributions to the pion cross sections for various processes as a function of the transverse momentum for bins in $z$ for positive pions in the thrust bin $0.85 < T < 0.9$. From bottom to top, the stacked contributions from $eec\bar{c}$ (yellow), $ees\bar{s}$ (dark blue), $eeu\bar{u}$ (purple), $\tau^+\tau^-$ (light green), $\Upsilon(4S)\rightarrow B^+B^-$ (violet), $\Upsilon(4S)\rightarrow B^0\bar{B}^0$ (dark green), charm (blue) and $uds$ (red) are shown. For comparison, the data for continuum (turquoise, denoted as ``data cont'') and on-resonance (orange, denoted as ``data res'') are also shown.}
\end{center}
\end{figure*}

\subsection{Non-$q \bar{q}$ background correction}
Several processes that are not part of the fragmentation function definitions need to be removed from the initial yields. These include the two-photon processes $e^+e^-\rightarrow e^+e^-u\bar{u}$, $e^+e^-\rightarrow e^+e^-d\bar{d}$, $e^+e^-\rightarrow e^+e^-s\bar{s}$, and $e^+e^-\rightarrow e^+e^-c\bar{c}$, as well as $\tau$ pair production and the $\Upsilon(4S)$ decays via either charged or neutral $B$ meson pairs. 
These contributions are extracted from MC and are directly subtracted from the luminosity-normalized yields. For all hadrons, the contributions from these processes are minor and only reach larger relative contributions in the higher transverse-momentum tails where two-photon processes and, for the pions also $\tau$ decays, contribute to more than 10\% of the yields. A large amount of $\Upsilon(4S)$ background has to be removed for low thrust values, and in particular at lower $z$ values, but at high thrust and high $z$ this contribution becomes negligible as the thrust variable very effectively discriminates against $\Upsilon(4S)$ decays.

Apart from the uncertainties due to the MC statistics used to determine these non-$q\bar{q}$ contributions, their relative sizes are also varied by $\pm 1.4~ \%$ for $\tau$ production \cite{taumc} and a factor of five for the two-photon contributions. The reason for this large factor in the two-photon contributions originates from the fact that not all possible diagrams are included in the MC generator. Those uncertainties are then assigned as systematic uncertainties for the non-$q\bar{q}$ removal.
The total relative background contributions for pions in an intermediate thrust bin, $0.85 < T < 0.9$, can be seen in Fig.~\ref{fig:nonqq:nonqqratio}. For kaons, the $\Upsilon(4S)$ decay contributions are even more pronounced at low thrust values and $z$, reaching initially more than 80\% of the yields before rapidly decreasing with $z$ and thrust value. For protons, the $\Upsilon(4S)$ contributions are again less dominant.
It should be noted that the large number of decays needed by $B$ mesons to produce the light hadrons studied here increases their contribution at higher transverse momenta disproportionately. Also the initial momentum of the $B$ mesons is small which enhances the possibility to find decay-hadrons at high transverse momenta.

\subsection{Momentum-smearing correction}
\begin{figure*}[ht]
\begin{center}
\includegraphics[width=0.8\textwidth]{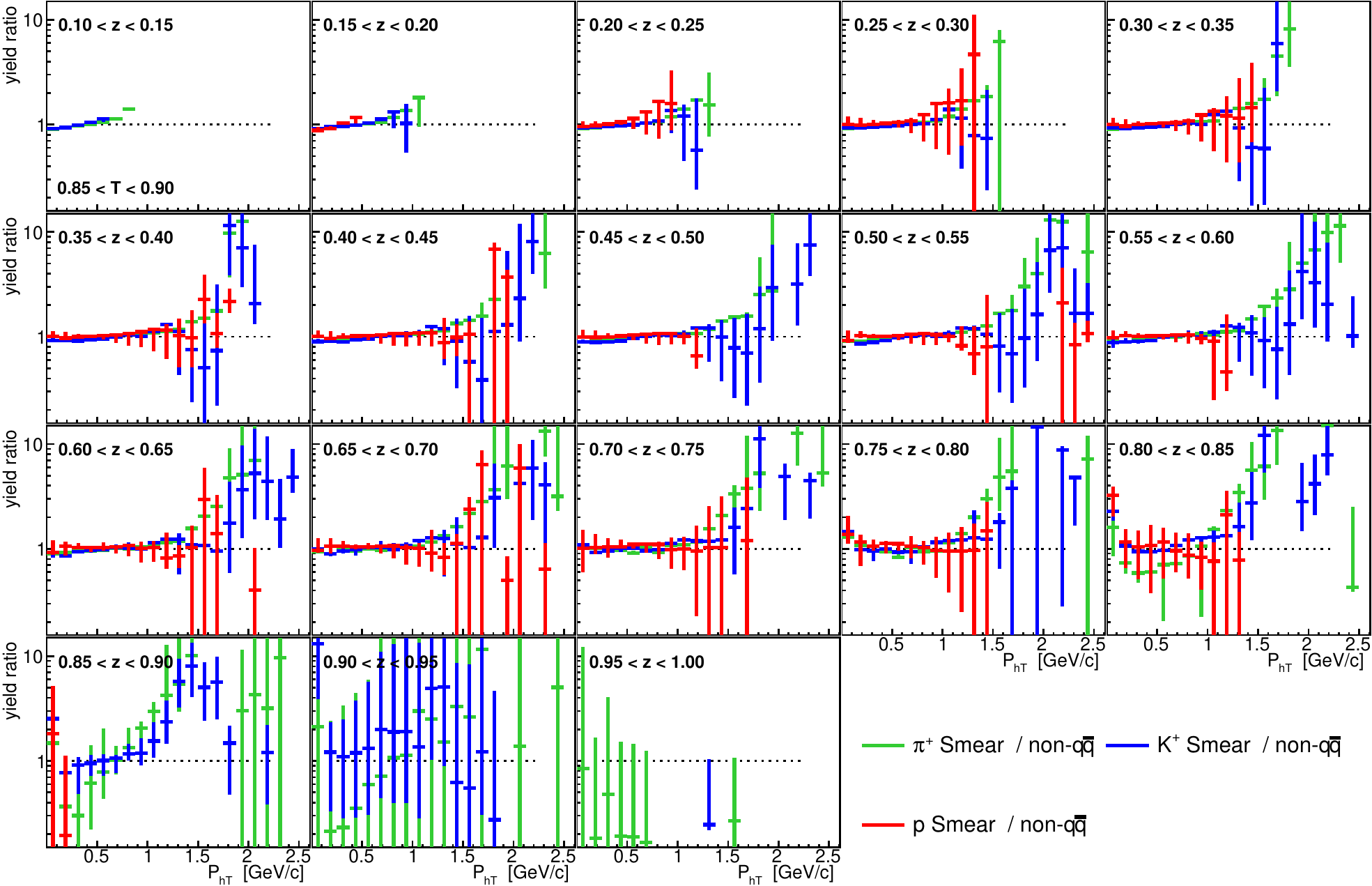} 
\caption{\label{fig:beforeaftersmearing}Ratio of yields after to before applying the smearing correction for positive pions (green), kaons (blue) and protons (red) as a function of $P_{hT}$ for the thrust bin $0.85 < T < 0.9$. Each panel corresponds to one $z$ bin for the corresponding hadrons. The error bars represent the systematic uncertainties assigned for this correction step.}
\end{center}
\end{figure*}

The momentum-smearing correction takes into account the momentum smearing in the detector as well as the smearing of the hadron transverse momentum due to the corresponding smearing of the thrust axis and its value. While the thrust axis is a good proxy for the initial quark-antiquark axis, the initial quark-antiquark axis itself is only meaningful in a leading-order picture. As such, we do not correct the axis smearing to the quark-antiquark axis but only to the true thrust axis based on all generated stable particle momenta. Examples of the thrust angular resolutions are shown in the supplement material.
The unfolding is performed using a singular-value decomposition technique \cite{Hocker:1995kb} as implemented in {\sc root} \cite{Brun:1997pa} taking into account only the $z\times P_{hT}\times T$ bins kinematically accessible. At small $z$, the diagonal elements dominate the matrices, but at higher $z$ the thrust algorithm biases the thrust axis closer to the high-$z$ hadron. The optimal regularization parameter for this unfolding is chosen as prescribed by the authors of the unfolding algorithm \cite{Hocker:1995kb}. As the determination of the correct regularization parameter is not simple for such large matrices, a variation of the rank parameter of up to 300 units is assigned as a systematic uncertainty. Additionally, the uncertainties due to the unfolding itself and the corresponding MC statistics are kept as systematic uncertainties. All previously extracted uncertainties are also unfolded.

 The final after-to-before ratio plots are displayed in Fig.~\ref{fig:beforeaftersmearing} as a function of $z$ and $P_{hT}$ bins, where one can see that predominantly the ratios are around unity at moderate $z$ and transverse momenta and increase for larger transverse momenta and higher $z$. The smearing correction is mostly similar for all particle types except for the larger transverse momentum tails where differences are visible. The behavior for other thrust ranges is similar.

\subsection{Preselection and acceptance correction}
\begin{figure*}[ht]
\begin{center}
\includegraphics[width=0.8\textwidth]{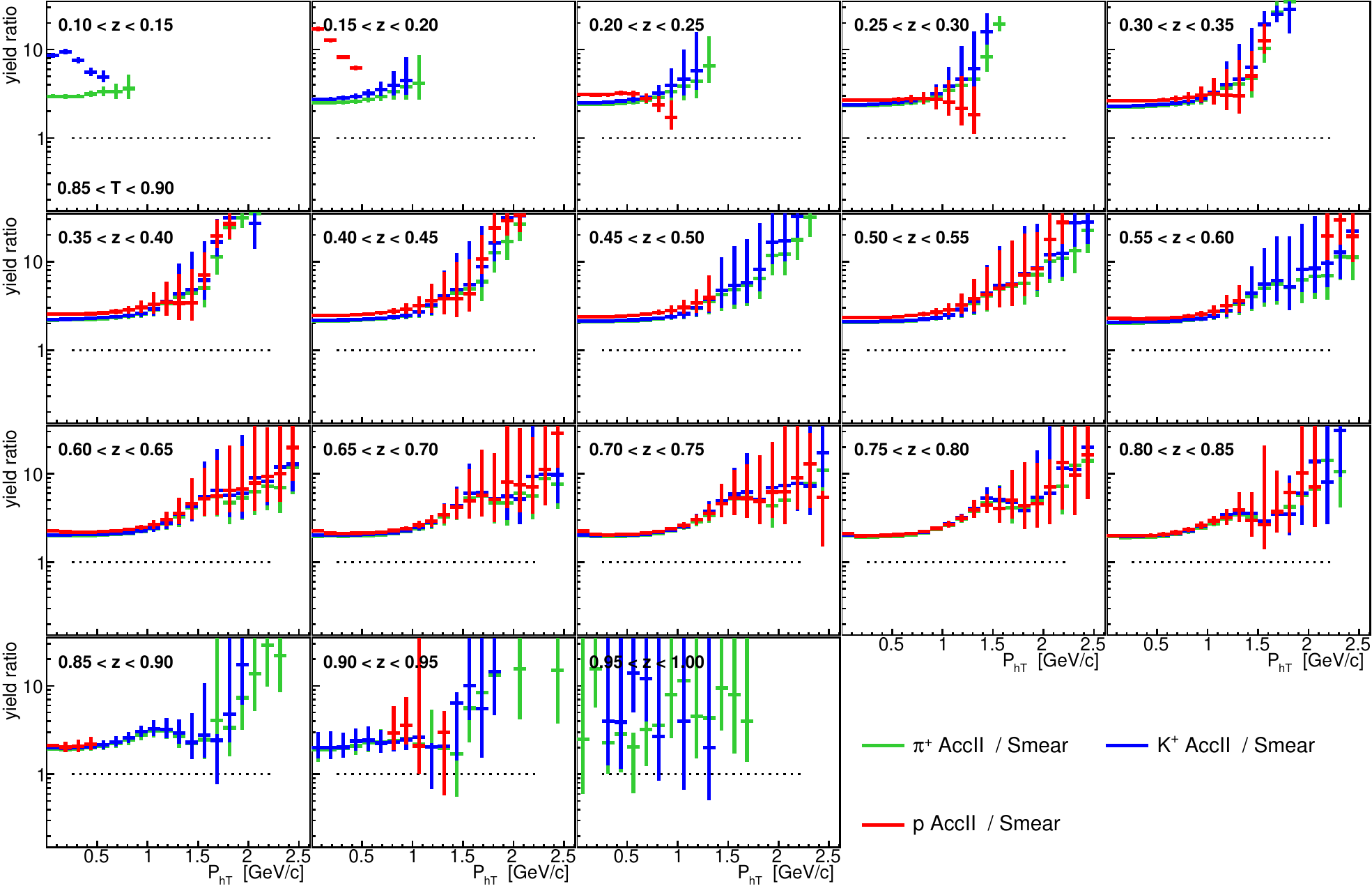} 
\caption{\label{fig:beforeafteraccepgtance}Ratio of yields after applying both acceptance corrections (denoted as AccII) to before (denoted as Smear) for positive pions (green), kaons (blue), and protons (red) for the thrust bin $0.85 < T < 0.9$. Each panel corresponds to one $z$ bin for the corresponding hadrons. The error bars represent the systematic uncertainties assigned for this correction step.}
\end{center}
\end{figure*}

As a next step, the reconstruction efficiencies and acceptance efficiencies are corrected for.
These corrections are performed in two steps to better understand the effects of the reconstruction and preselection, and the effects purely due to using only the barrel acceptance for this analysis. At this correction stage, the restriction of the thrust axis pointing into the central area of the detector is lifted as well as the 100 MeV/c minimal laboratory transverse momentum requirement for tracks.
Within the barrel acceptance, the corrections are generally moderate and only increase slightly with increasing transverse momentum, except for high fractional energies where the corrections appear to be falling slightly with increasing $P_{hT}$. With the exception of protons at lower $z$ close to the mass threshold, pions, kaons, and protons behave very similarly.
The corresponding figures for the ratios of yields after acceptance corrections can be seen in Fig.~\ref{fig:beforeafteraccepgtance} for pions, kaons, and protons. \par

Correcting from the barrel acceptance to the full acceptance affects higher transverse momenta more, where the angle between the thrust axis and the hadron gets large and thus also the possibility to miss the barrel. Pions, kaons, and protons again behave similarly with the exception of the very lowest $z$ bins, where the differences in actual momentum are relevant. For both correction steps, the statistical uncertainties of correction factors from the MC samples are included as systematic uncertainties. In the second acceptance correction, the variation of the correction factor with fragmentation tunes (JETSET settings optimized for various collision systems and energies) is included as systematic uncertainties. As the latter would move the central values up or down for all bins, they are considered correlated.

\subsection{Weak decays}
Weak decays are in principle not part of the fragmentation function definitions and might alter the applicability of DGLAP \cite{Gribov:1972ri,Dokshitzer:1977sg,Altarelli:1977zs} or other evolution schemes. However, in practice, many weak decays cannot be experimentally removed and one has to rely on information from MC.
Because of this additional uncertainty, weak decays are often no longer removed from reported fragmentation results. In this analysis we provide cross sections either containing all weak decays or removing all of them based on MC. The relative fraction of weak decays are different for the various hadron types. For pions, light-quark production via strong decays dominates, while charm decays provide pions mostly through weak decays. These contributions are not flat in transverse momentum and weak decays at intermediate $z$ have a larger contribution at intermediate transverse momenta around 1 GeV/c. In contrast, at even higher transverse momenta, strong charm decays provide a sizable contribution.
For kaons, the transverse momentum and $z$ behavior is similar but the overall fraction of weak decays is larger due to the preferred charm decays into kaons.
For protons, charm decays are generally less pronounced while hyperon decays of light (including strange) quarks provide many weak-decay channels.
For the weak-decay corrected results, systematic uncertainties for this correction are assigned based on the statistical uncertainties from the MC and the variation of the {\sc pythia} fragmentation tune. 

\subsection{ISR correction}
\begin{figure*}[t]
\begin{center}
\includegraphics[width=0.8\textwidth]{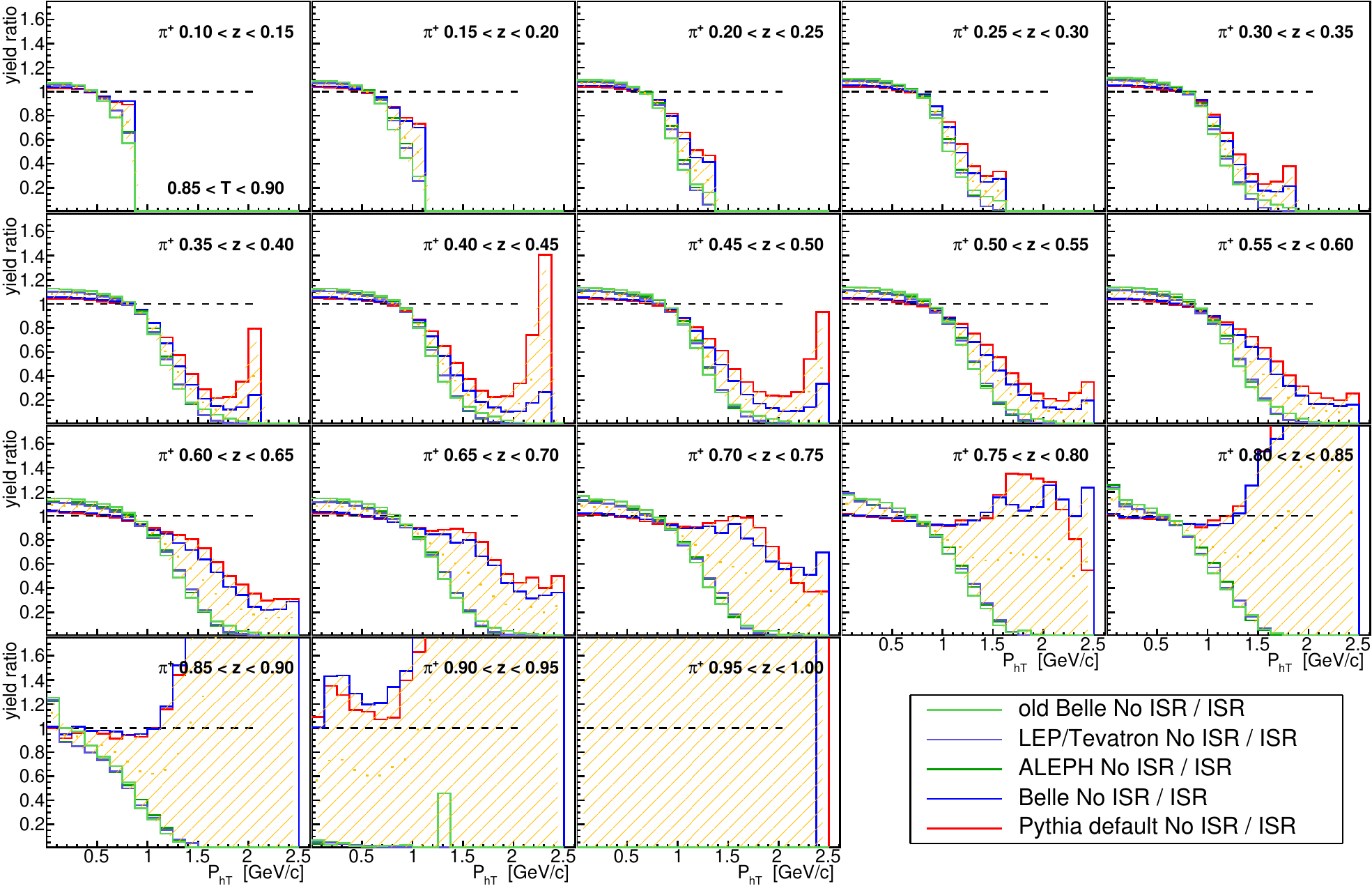}
\caption{\label{fig:isrzfraction_pid0_mix0}Non-ISR over ISR cross section ratios as a function of $z$ and $P_{hT}$ for positive pions in the thrust bin $0.85 < T < 0.9$ for various MC tunes. The yellow, hatched regions display the variation of these ratios with tunes and are assigned as systematic uncertainties. }
\end{center}
\end{figure*}

Initial-state radiation (ISR) is treated similarly to the previous publication \cite{Seidl:2017qhp} by comparing the cross section in MC with and without ISR effects. As shown in Fig.~\ref{fig:isrzfraction_pid0_mix0}, at low transverse momenta, these effects are very minor and only the accessibility of higher $z$ events creates increasingly larger non-ISR cross sections. However, ISR changes the boost of the hadronic system and its thrust axis, and therefore accumulates hadrons at larger transverse momenta which produces a very substantial fraction of events at high transverse momenta. The origin of this increase is confirmed by explicitly reconstructing the boosted quark-antiquark system in the presence of ISR photons and comparing it to the nominal transverse momentum.
Due to relying on MC to correct for ISR effects, the variation between several fragmentation tunes is taken as an additional systematic uncertainty and is in many bins the dominant systematic uncertainty. The central value of the ISR correction for various tunes has been used for the correction. Especially at the higher $z$ bins, the differences between tunes become substantial in the transverse momentum tails. These uncertainties again affect all bins and are treated as correlated uncertainties.
The statistical uncertainties of the MC sample used to calculate these ratios are also assigned as uncorrelated systematic uncertainties.

\subsection{Consistency checks and total systematic uncertainties\label{sec:syst}}
\begin{figure*}[htb]
\begin{center}
\includegraphics[width=0.85\textwidth]{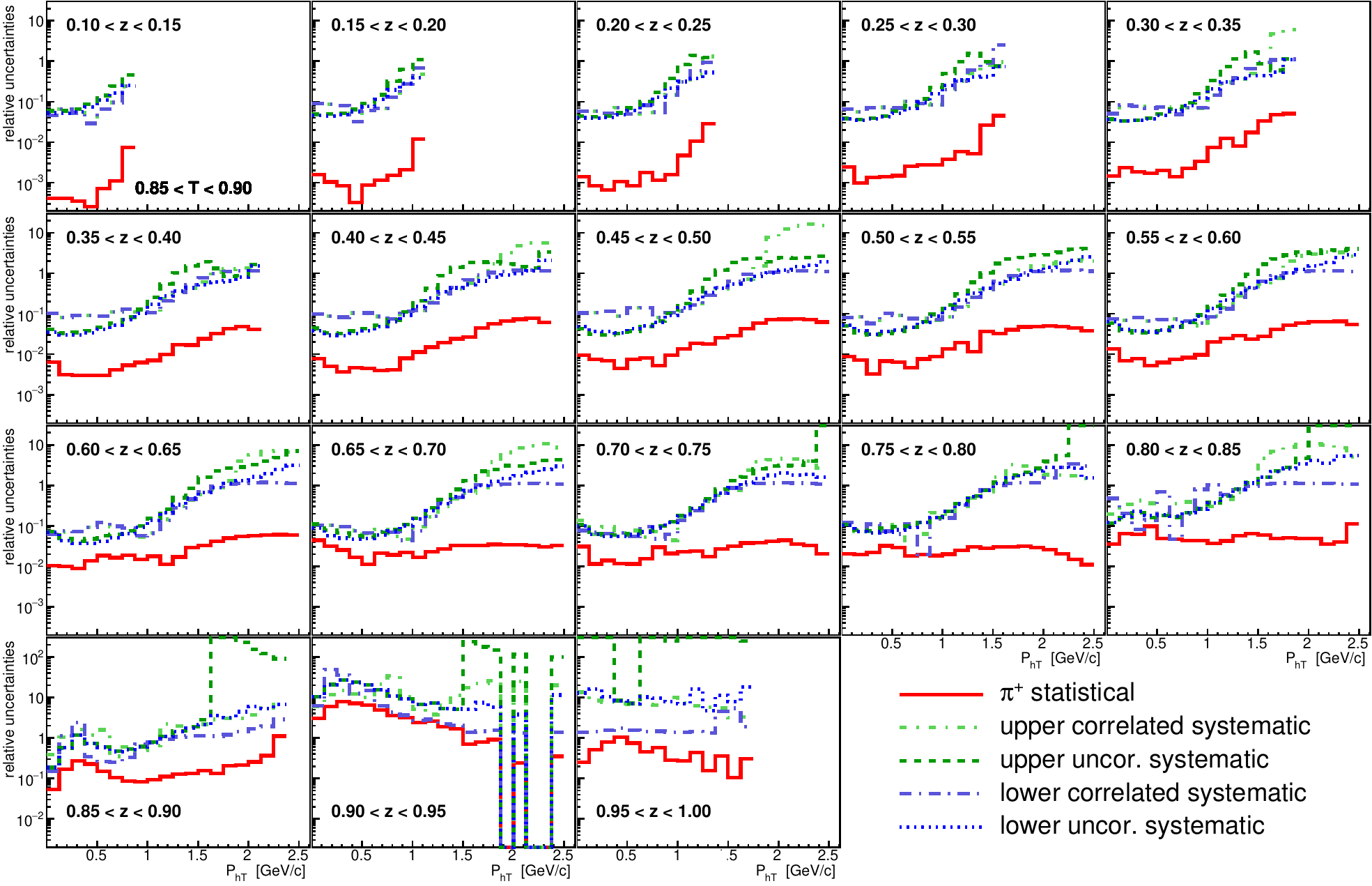}
\caption{\label{fig:systall_mix0} Relative (asymmetric) correlated (lower uncertainties: dashed-dotted blue lines; upper uncertainties: dash-dotted dark-green lines) and uncorrelated (lower uncertainties: dotted purple lines; upper uncertainties: dashed green lines) systematic and statistical uncertainties (full lines) for pions as a function of $P_{hT}$ in bins of $z$ in the thrust bin $0.85<T<0.9$.}
\end{center}
\end{figure*}

In addition to the systematic uncertainties based on the various corrections previously described, there are two global uncertainties due to the luminosity evaluation (1.4\%) and the track reconstruction (0.35\%). The overall bin-by-bin systematic uncertainties are summarized in Fig.~\ref{fig:systall_mix0}. It can be seen that generally the systematic uncertainties dominate and all systematics increase with increasing transverse momentum. The systematic uncertainties in turn are generally dominated by the uncertainties from the variation of {\sc pythia} fragmentation tunes in the various correction steps. The uncertainties from the smearing correction are also substantial, particularly at larger transverse momenta. The bin-by-bin systematic uncertainties are kept separate for all correlated and uncorrelated uncertainties, as those need to be treated differently when using the cross-section data to extract the widths of Gaussian fits to the $P_{hT}$ dependence or perform global fragmentation fits. In the cross-section result figures of the next section, they are however displayed as the quadratic sum.

Before combining opposite charges, their consistency has been confirmed. Also a second data set taken below the $\Upsilon(4S)$ resonance is found to be consistent such that no additional systematic uncertainties need to be assigned. The consistency with the previously published transverse-momentum and thrust-integrated cross sections \cite{Seidl:2015lla} is confirmed after applying the same ISR correction method to the previous analysis.

\section{Results}

\begin{figure*}[htb]
\begin{center}
\includegraphics[width=0.9\textwidth]{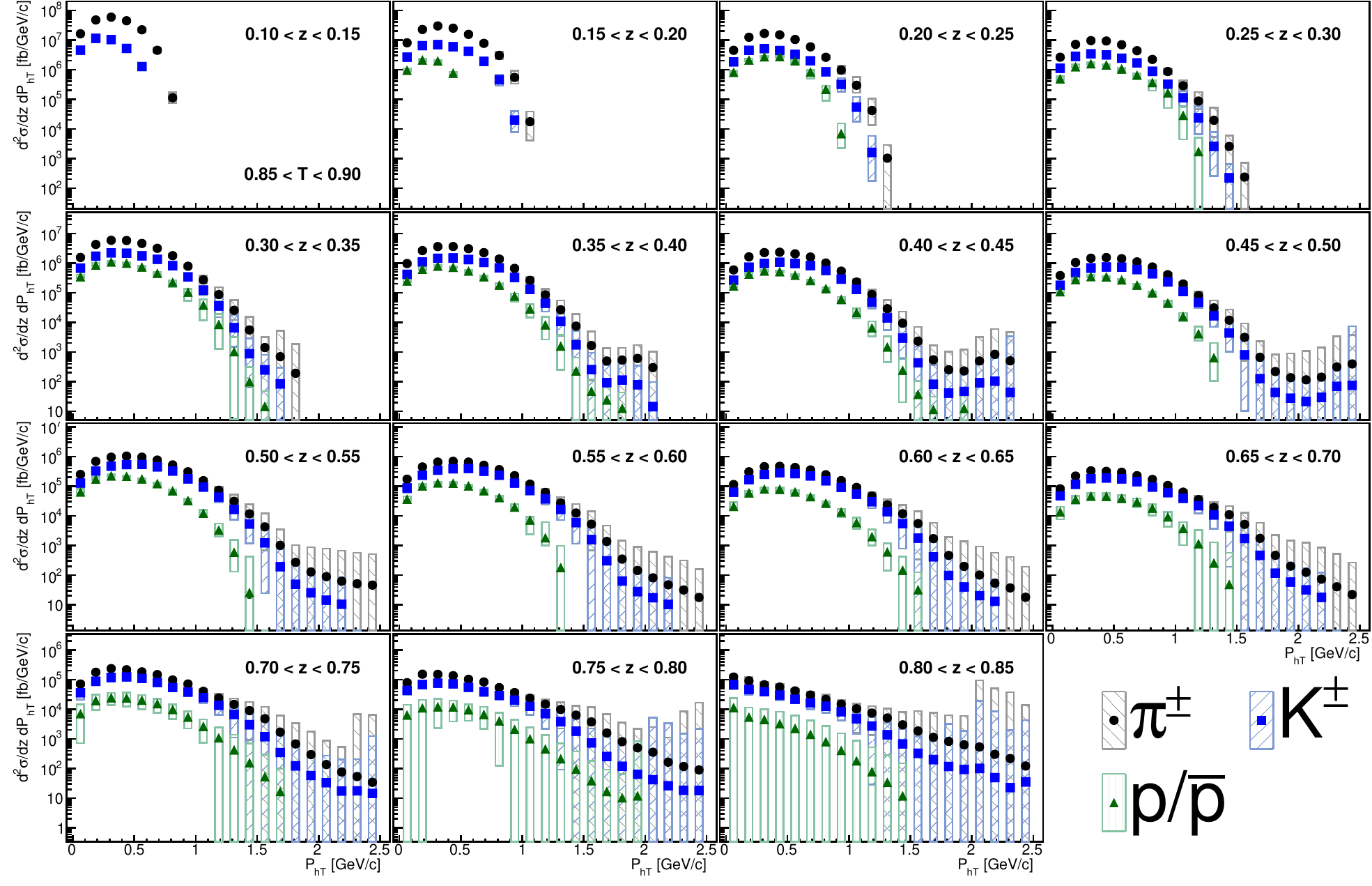}

\caption{\label{fig:allxsec_mix0_pid21}Differential cross sections for pions (black circles), kaons (blue squares), and protons (green triangles) as a function of $P_{hT}$ for the indicated $z$ bins and thrust $0.85 <T <0.9$. The error boxes represent the systematic uncertainties. Due to the large uncertainties in them, $z$ bins above 0.85 are not displayed.}
\end{center}
\end{figure*}

The differential cross sections for pions, kaons, and protons as a function of fractional energy and transverse momentum are given in Fig.~\ref{fig:allxsec_mix0_pid21} for an intermediate thrust bin. Due to the large uncertainties in them, $z$ bins above 0.85 are not displayed. The behavior is generally quite similar between hadron types. Only the proton cross sections have a steeper transverse momentum dependence than the light mesons. At lower transverse momenta, the behavior resembles a Gaussian which is also generally assumed for TMD fragmentation functions, while the tails extend further than the Gaussians. However, the systematic uncertainties in the tails are generally too large to conclude much about their behavior.

\subsection{Widths of $P_{hT}$ dependence}
\begin{figure*}[htb]
\begin{center}
\includegraphics[width=0.9\textwidth]{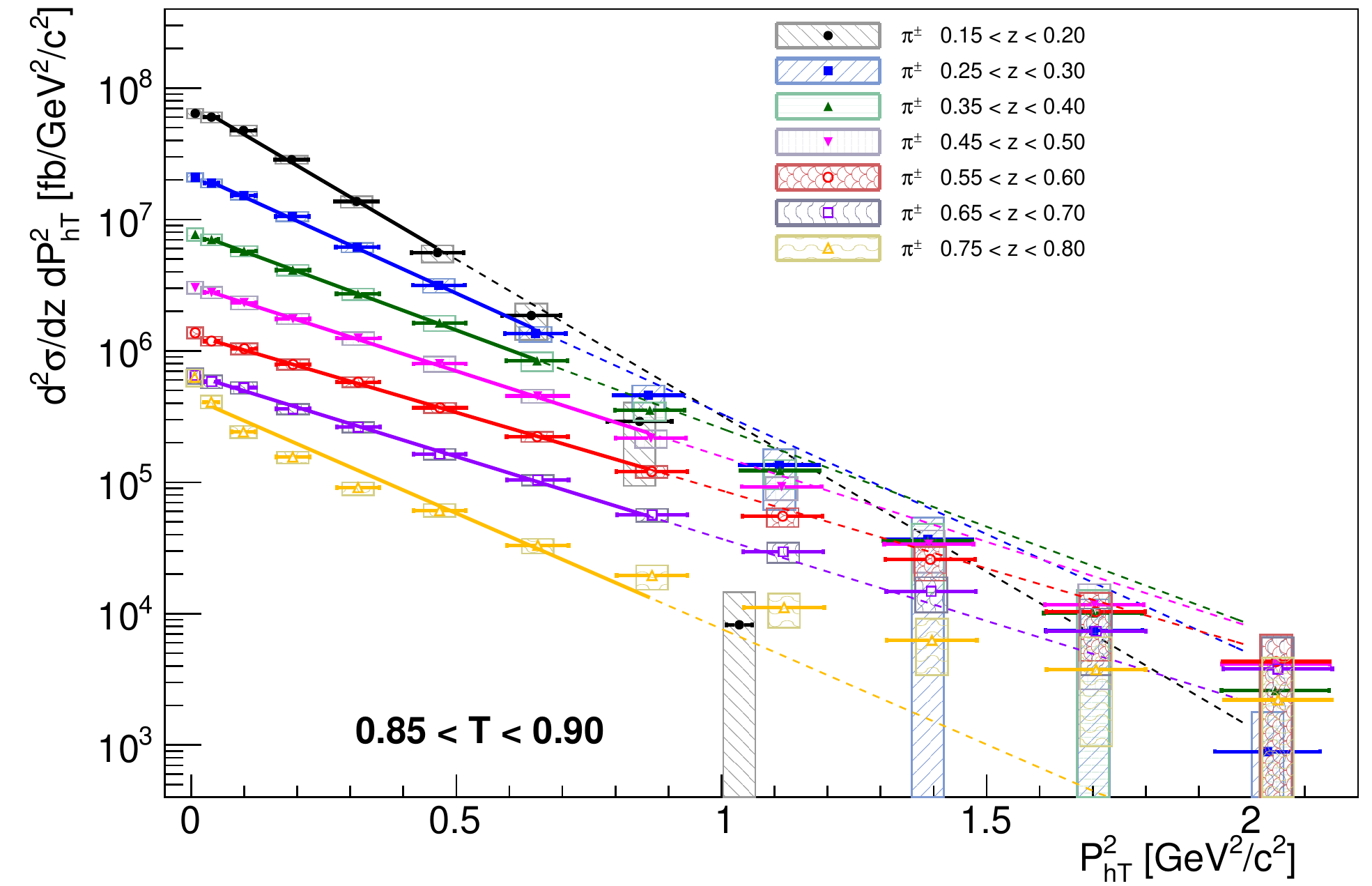}

\caption{\label{fig:sktcomp}Single charged pion cross sections as a function of $P_{hT}^2$ for selected bins of fractional energy $z$ and thrust $0.85 <T <0.9$. The full lines at lower transverse momenta correspond to the Gaussian fits to this data using the same color coding as for the data. They are extended as dotted lines to larger transverse momenta not included in the fit. Each datapoint is displayed at the bin's central value while horizontal uncertainties display the RMS value. The error boxes represent the systematic uncertainties.}
\end{center}
\end{figure*}

Motivated by the dependence in the lower transverse momentum region, a Gaussian dependence is fit to the data in this region as can be seen in Fig.~\ref{fig:sktcomp} for pions as a function of $P_{hT}^2$ (so the Gaussian distribution appears as a straight line on a single logarithmic plot). The statistical uncertainties are included in this fit while the systematic uncertainties are assessed by varying the individual contributions according to their own uncertainties using a sampling method with 70000 iterations. The correlations between transverse momentum bins are taken into account for the corresponding uncertainties since they are separated into fully correlated and uncorrelated contributions. At present the correlations are only considered between $P_{hT}$ bins, not between $z$ or thrust bins. It is expected that while global variations of the latter could have a substantial impact on the absolute magnitude of the cross sections, the shape of the Gaussian behavior is less affected. After each uncertainty-sampling iteration, the Gaussian width is fit again. The mean value and 32 percentile ranges above and below the mean are determined from the distributions over all iterations. Their differences to the main fit are then assigned as upper and lower systematic uncertainties on the Gaussian widths. The inclusion of correlated systematic uncertainties between different $P_{hT}$ bins generally reduces the uncertainties on the Gaussian widths as the variation of points moves all points similarly up or down while the shape stays more robust. The range in $P_{hT}$ used in the Gaussian fit is varied and the resulting variation of the fit results is assigned as an additional systematic uncertainty on the Gaussian widths. At intermediate fractional energies, the fit ranges were motivated by the non-perturbative range of transverse momenta below 1 GeV. However, in particular at very small and large fractional energies, the number of points becomes limited, resulting in very large uncertainties on these widths. Those $z$ bins for which either the main fit fails or the uncertainties on the widths covers zero or unity are not shown.\par
\begin{figure*}[htb]
\begin{center}
\includegraphics[width=0.9\textwidth]{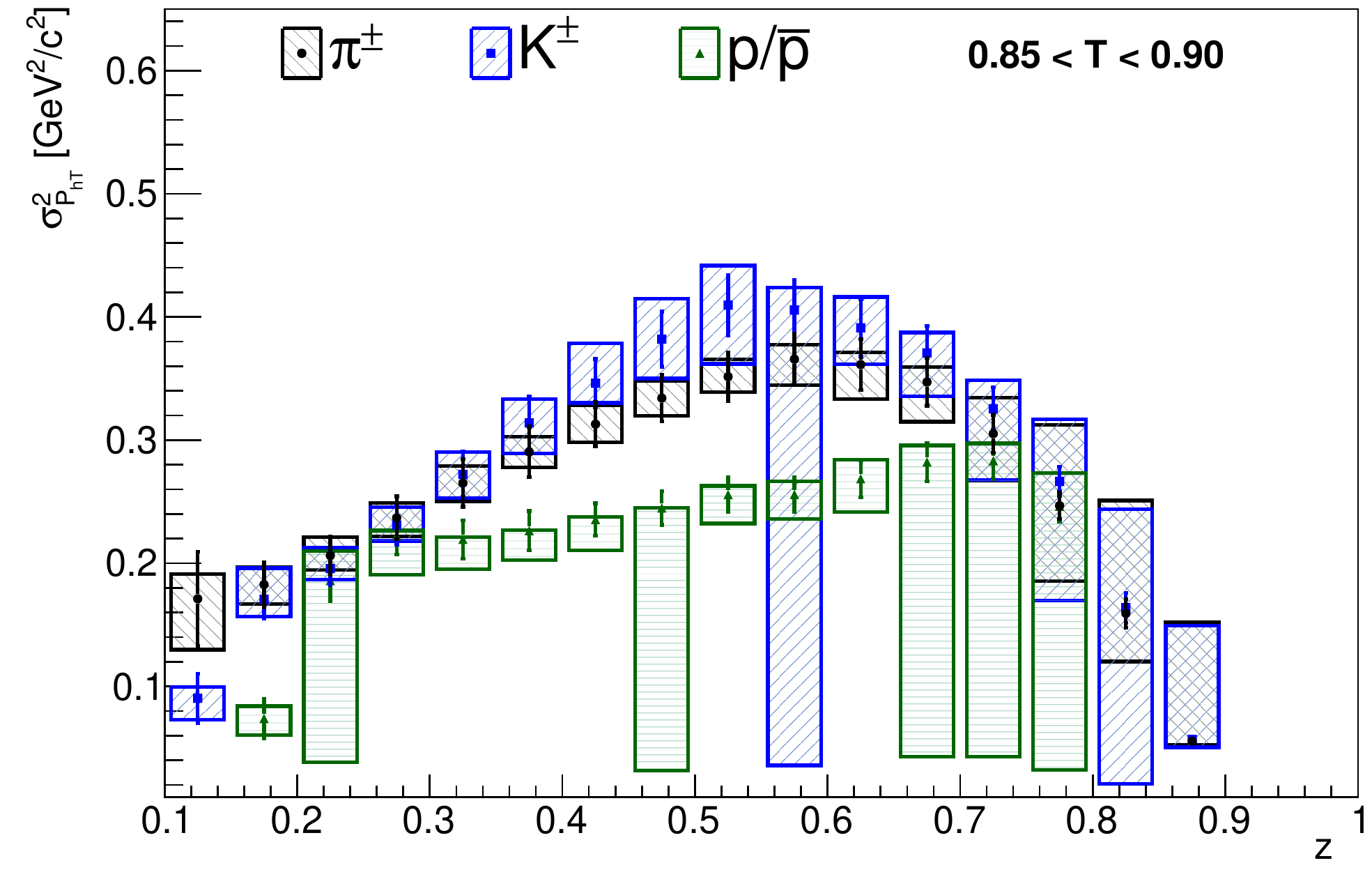}

\caption{\label{fig:sktfit2}Gaussian widths as a function of $z$ for pions (black circles and boxes), kaons (blue squares and boxes), and protons (green triangles and boxes) and thrust $0.85 <T <0.9$. The error boxes represent the corresponding systematic uncertainties as described in the text.}
\end{center}
\end{figure*}

The extracted Gaussian variances for pions, kaons, and protons are then summarized in Fig.~\ref{fig:sktfit2} as a function of $z$ in the $0.85< T < 0.9$ thrust bin. At very low $z$ the phase space is very limited and consequently the Gaussian widths are not very well constrained. A similar behavior is also relevant at very large fractional energies. In the intermediate $z$ range, the widths are mostly similar for pions and kaons but those for pions are generally slightly smaller. Protons, however, have substantially narrower widths at intermediate fractional energies while being closer to pions and kaons at low and high $z$. This might point to the differences in production mechanisms for mesons and baryons, as for the latter the production of di-quark pairs is additionally needed. In all cases the Gaussian widths are increasing with $z$ until around 0.6 before decreasing again at higher fractional energies.

\begin{figure*}[htb]
\begin{center}
\includegraphics[width=0.9\textwidth]{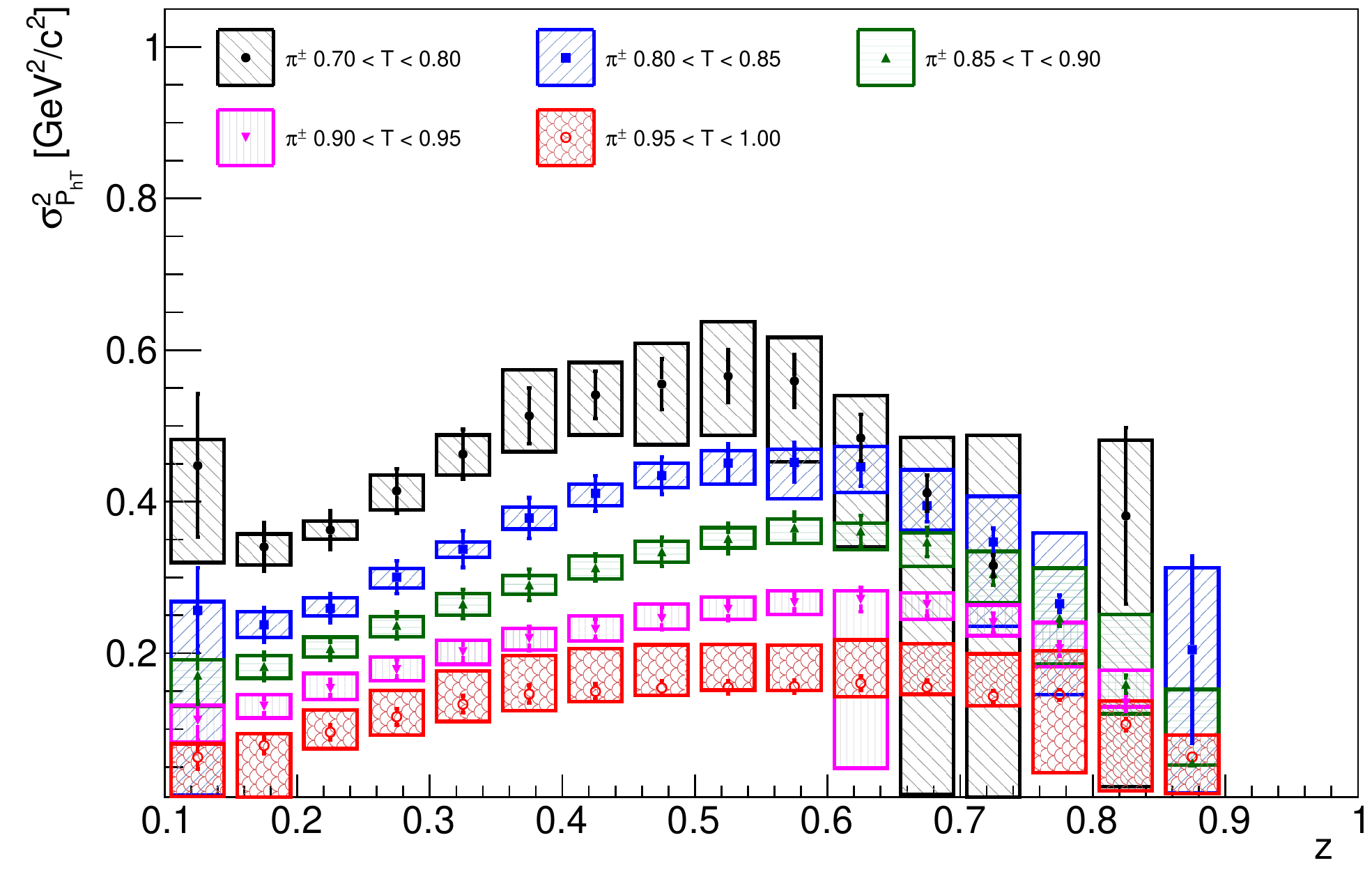}

\caption{\label{fig:sktfitthrust}Gaussian widths as a function of $z$ for pions and thrust $0.7 <T <0.8$ (black circles and boxes), thrust $0.8 <T <0.85$ (blue squared and boxes), thrust $0.85 <T <0.9$ (green triangles and boxes), thrust $0.9 <T <0.95$ (magenta triangles and boxes) and $0.95 <T <1.0$ (red circles and boxes). The error boxes represent the corresponding systematic uncertainties.}
\end{center}
\end{figure*}

\begin{figure*}[htb]
\begin{center}
\includegraphics[width=0.8\textwidth]{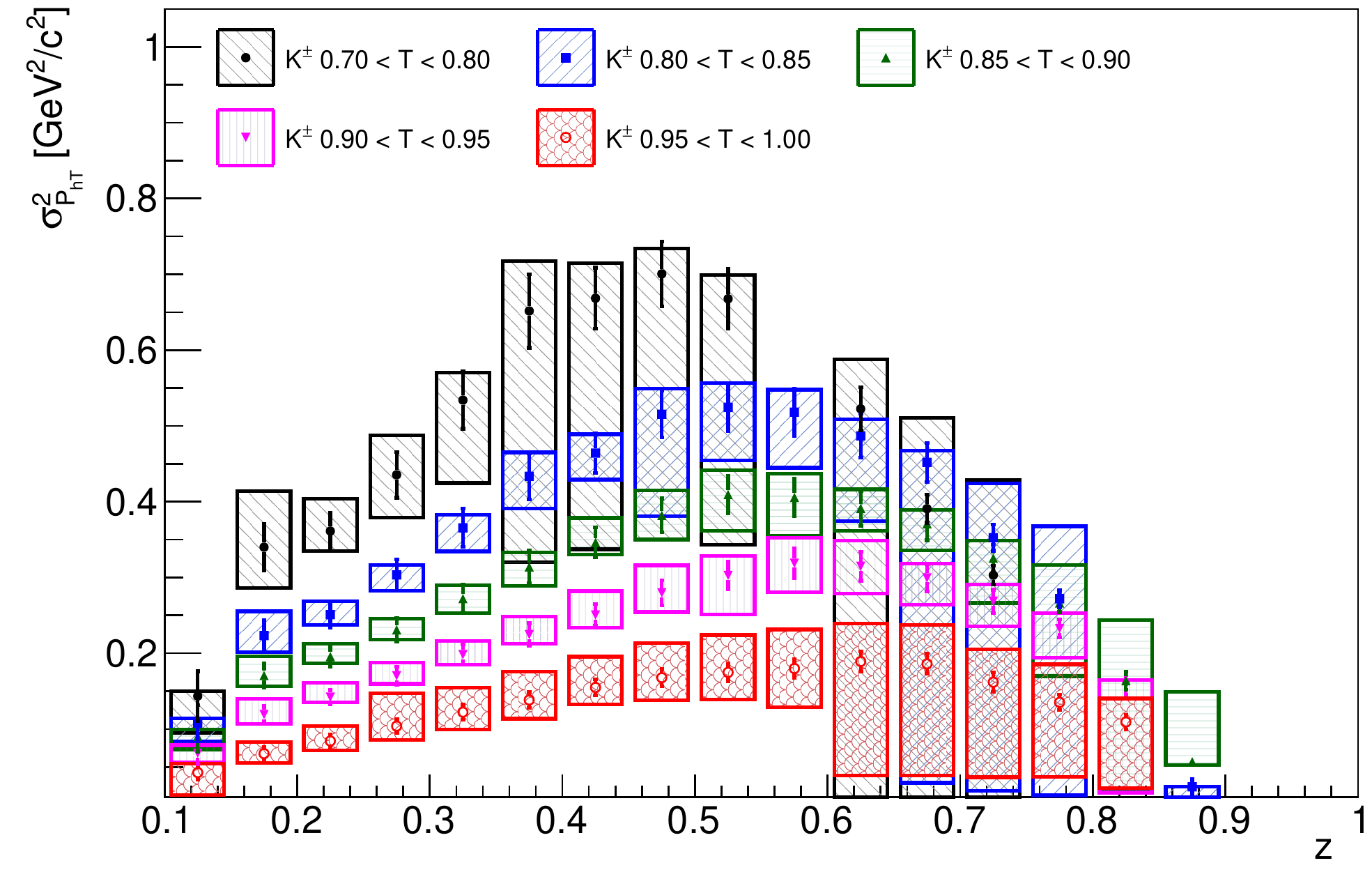}

\caption{\label{fig:sktfitthrust_k}Gaussian widths as a function of $z$ for kaons and thrust $0.7 <T <0.8$ (black circles and boxes), thrust $0.8 <T <0.85$ (blue squared and boxes), thrust $0.85 <T <0.9$ (green triangles and boxes), thrust $0.9 <T <0.95$ (magenta triangles and boxes) and $0.95 <T <1.0$ (red circles and boxes). The error boxes represent the corresponding systematic uncertainties.}
\end{center}
\end{figure*}

\begin{figure*}[htb]
\begin{center}
\includegraphics[width=0.8\textwidth]{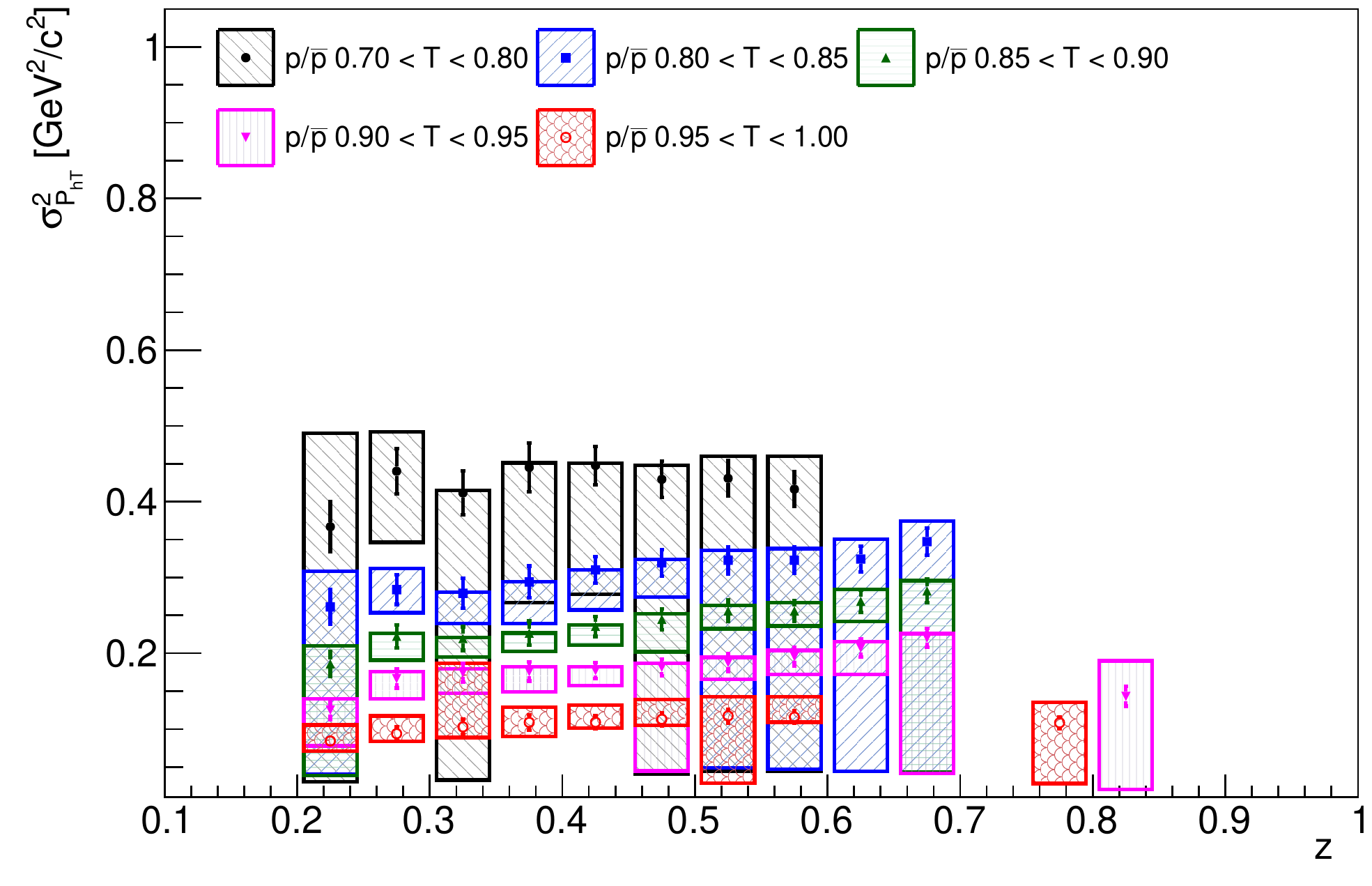}

\caption{\label{fig:sktfitthrust_p}Gaussian widths as a function of $z$ for protons and thrust $0.7 <T <0.8$ (black circles and boxes), thrust $0.8 <T <0.85$ (blue squared and boxes), thrust $0.85 <T <0.9$ (green triangles and boxes), thrust $0.9 <T <0.95$ (magenta triangles and boxes) and $0.95 <T <1.0$ (red circles and boxes). The error boxes represent the corresponding systematic uncertainties.}
\end{center}
\end{figure*}

It is also interesting to study the behavior of the Gaussian widths for the different thrust ranges. These are shown for pions in Fig.~\ref{fig:sktfitthrust}. At very low thrust, any reference direction is as good as any other, resulting in a nearly flat distribution of transverse momenta. Consequently, the Gaussian widths cannot be well extracted or become very large. For all other thrust ranges, the widths show the same general behavior: increasing toward intermediate $z$ before decreasing again. They are ordered according to the thrust ranges with the lowest thrust having the largest widths and vice versa. This correlation can be understood by the high-thrust limit, where the event is very collimated along the thrust axis and therefore little transverse momentum with respect to this axis is available.
The behavior of the Gaussian widths for different thrust bin values is also shown for kaons and protons in Figs.~\ref{fig:sktfitthrust_k} and \ref{fig:sktfitthrust_p}, respectively. For kaons, the same narrowing of the widths with increasing thrust can be seen as observed for the pions. Also for the protons, the thrust dependence is similar but the uncertainties start to overlap in many $z$ bins, making the effect less pronounced.

\subsection{MC generator comparison}
\begin{figure*}[htb]
\begin{center}
\includegraphics[width=0.9\textwidth]{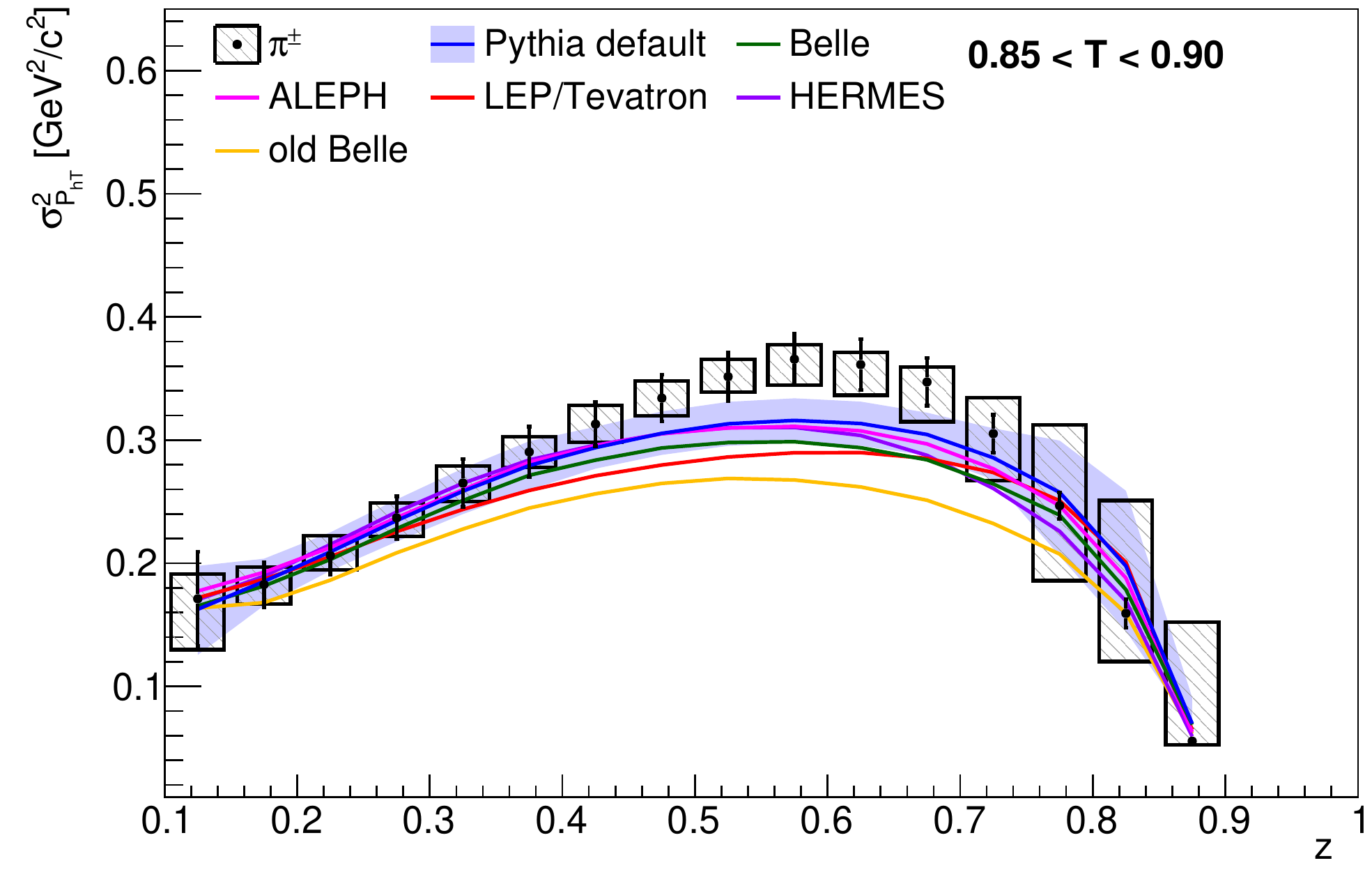}

\caption{\label{fig:sktfitmc}Gaussian widths as a function of $z$ for charged pions (black circles and boxes) in comparison to various {\sc pythia} tunes. The error boxes represent the corresponding systematic uncertainties. Fit statistical and range uncertainties for the MC samples are representatively visualized for the {\sc pythia} default setting only as the filled area. }
\end{center}
\end{figure*}

One can study the behavior of various {\sc pythia} tunes on the transverse momentum dependence. It should be noted that the overall $z$ dependence has already been discussed in previous publications \cite{Seidl:2015lla,Seidl:2017qhp}, showing that only a few tunes are reasonably close to the actual data, while others either largely over or undershoot them, particularly at high $z$. The Gaussian widths, however, are not sensitive to either the $z$ behavior nor the overall normalization. In {\sc pythia} they are very directly related to the variable ParJ(21), which ranges between 0.28 and 0.4 in these tunes and describes the Gaussian widths for primary hadrons within the LUND string model\cite{Sjostrand:1993yb}. The Gaussian widths are partially also sensitive to the variable ParJ(42), which ranges from 0.54 to 0.80 and describes the inverse of the width of the transverse mass in the LUND string model. With the exception of the old Belle tune (ParJ(21)=0.28), all tunes have very similar Gaussian widths and reproduce both the small and larger fractional energies well. At intermediate $z$, the {\sc pythia} default tune and the tunes with larger ParJ(21) seem to get closest to the data but fail to fully describe the maximum widths. The comparison for intermediate thrust values can be seen in Fig.~\ref{fig:sktfitmc}.

The individual pion, kaon and proton cross sections as a function of fractional energy, thrust value and transverse momentum as well as the extracted Gaussian widths are provided online in the supplemental files \cite{supplement} together with cross section and Gaussian width figures for other thrust bins. 
\section{Summary}
We provide the first direct transverse-momentum-dependent single-hadron production cross sections in $e^+e^-$ collisions at $\sqrt{s} = 10.58$ GeV for pions, kaons, and protons as a function of fractional energy $z$ and the thrust value. In addition, it is found that a Gaussian functional form describes well the transverse-momentum dependence at small transverse momenta. The Gaussian widths vary with $z$ and thrust. This data will help to understand the intrinsic transverse momentum dependence generated in the fragmentation process. Such input is needed to obtain a better theoretical description of the various transverse-momentum-dependent and related higher-twist effects visible in transverse spin asymmetries in semi-inclusive deep inelastic scattering, proton-proton collisions and electron-positron annihilation. This information also leads the way toward high-precision measurements of TMD effects at the electron-ion collider. In addition, these results provide the unpolarized baseline for any polarized, transverse-momentum-dependent fragmentation functions such as the Collins FF.

\begin{acknowledgments}
We thank the KEKB group for the excellent operation of the
accelerator; the KEK cryogenics group for the efficient
operation of the solenoid; and the KEK computer group, and the Pacific Northwest National
Laboratory (PNNL) Environmental Molecular Sciences Laboratory (EMSL)
computing group for strong computing support; and the National
Institute of Informatics, and Science Information NETwork 5 (SINET5) for
valuable network support.  We acknowledge support from
the Ministry of Education, Culture, Sports, Science, and
Technology (MEXT) of Japan, the Japan Society for the 
Promotion of Science (JSPS), and the Tau-Lepton Physics 
Research Center of Nagoya University; 
the Australian Research Council including grants
DP180102629, 
DP170102389, 
DP170102204, 
DP150103061, 
FT130100303; 
Austrian Science Fund under Grant No.~P 26794-N20;
the National Natural Science Foundation of China under Contracts
No.~11435013,  
No.~11475187,  
No.~11521505,  
No.~11575017,  
No.~11675166,  
No.~11705209;  
Key Research Program of Frontier Sciences, Chinese Academy of Sciences (CAS), Grant No.~QYZDJ-SSW-SLH011; 
the  CAS Center for Excellence in Particle Physics (CCEPP); 
the Shanghai Pujiang Program under Grant No.~18PJ1401000;  
the Ministry of Education, Youth and Sports of the Czech
Republic under Contract No.~LTT17020;
the Carl Zeiss Foundation, the Deutsche Forschungsgemeinschaft, the
Excellence Cluster Universe, and the VolkswagenStiftung;
the Department of Science and Technology of India; 
the Istituto Nazionale di Fisica Nucleare of Italy; 
National Research Foundation (NRF) of Korea Grants
No.~2015H1A2A1033649, No.~2016R1D1A1B01010135, No.~2016K1A3A7A09005
603, No.~2016R1D1A1B02012900, No.~2018R1A2B3003 643,
No.~2018R1A6A1A06024970, No.~2018R1D1 A1B07047294; Radiation Science Research Institute, Foreign Large-size Research Facility Application Supporting project, the Global Science Experimental Data Hub Center of the Korea Institute of Science and Technology Information and KREONET/GLORIAD;
the Polish Ministry of Science and Higher Education and 
the National Science Center;
the Grant of the Russian Federation Government, Agreement No.~14.W03.31.0026; 
the Slovenian Research Agency;
Ikerbasque, Basque Foundation for Science, Spain;
the Swiss National Science Foundation; 
the Ministry of Education and the Ministry of Science and Technology of Taiwan;
and the United States Department of Energy and the National Science Foundation.
We also thank Zhongbo Kang for very fruitful discussions which led to the inclusion of the thrust binning in this analysis. 
\end{acknowledgments}

\bibliography{singlehadkt_v1.3}

\end{document}

%% file: pub531.tex
\noaffiliation
\affiliation{University of the Basque Country UPV/EHU, 48080 Bilbao}
\affiliation{Beihang University, Beijing 100191}
\affiliation{Brookhaven National Laboratory, Upton, New York 11973}
\affiliation{Budker Institute of Nuclear Physics SB RAS, Novosibirsk 630090}
\affiliation{Faculty of Mathematics and Physics, Charles University, 121 16 Prague}
\affiliation{Chonnam National University, Kwangju 660-701}
\affiliation{University of Cincinnati, Cincinnati, Ohio 45221}
\affiliation{Deutsches Elektronen--Synchrotron, 22607 Hamburg}
\affiliation{Duke University, Durham, North Carolina 27708}
\affiliation{University of Florida, Gainesville, Florida 32611}
\affiliation{Key Laboratory of Nuclear Physics and Ion-beam Application (MOE) and Institute of Modern Physics, Fudan University, Shanghai 200443}
\affiliation{Justus-Liebig-Universit\"at Gie\ss{}en, 35392 Gie\ss{}en}
\affiliation{Gifu University, Gifu 501-1193}
\affiliation{II. Physikalisches Institut, Georg-August-Universit\"at G\"ottingen, 37073 G\"ottingen}
\affiliation{SOKENDAI (The Graduate University for Advanced Studies), Hayama 240-0193}
\affiliation{Hanyang University, Seoul 133-791}
\affiliation{University of Hawaii, Honolulu, Hawaii 96822}
\affiliation{High Energy Accelerator Research Organization (KEK), Tsukuba 305-0801}
\affiliation{J-PARC Branch, KEK Theory Center, High Energy Accelerator Research Organization (KEK), Tsukuba 305-0801}
\affiliation{Forschungszentrum J\"{u}lich, 52425 J\"{u}lich}
\affiliation{IKERBASQUE, Basque Foundation for Science, 48013 Bilbao}
\affiliation{Indian Institute of Science Education and Research Mohali, SAS Nagar, 140306}
\affiliation{Indian Institute of Technology Guwahati, Assam 781039}
\affiliation{Indian Institute of Technology Hyderabad, Telangana 502285}
\affiliation{Indian Institute of Technology Madras, Chennai 600036}
\affiliation{Indiana University, Bloomington, Indiana 47408}
\affiliation{Institute of High Energy Physics, Chinese Academy of Sciences, Beijing 100049}
\affiliation{Institute of High Energy Physics, Vienna 1050}
\affiliation{Institute for High Energy Physics, Protvino 142281}
\affiliation{INFN - Sezione di Napoli, 80126 Napoli}
\affiliation{INFN - Sezione di Torino, 10125 Torino}
\affiliation{Advanced Science Research Center, Japan Atomic Energy Agency, Naka 319-1195}
\affiliation{J. Stefan Institute, 1000 Ljubljana}
\affiliation{Institut f\"ur Experimentelle Teilchenphysik, Karlsruher Institut f\"ur Technologie, 76131 Karlsruhe}
\affiliation{Kennesaw State University, Kennesaw, Georgia 30144}
\affiliation{Kitasato University, Sagamihara 252-0373}
\affiliation{Korea Institute of Science and Technology Information, Daejeon 305-806}
\affiliation{Korea University, Seoul 136-713}
\affiliation{Kyungpook National University, Daegu 702-701}
\affiliation{LAL, Univ. Paris-Sud, CNRS/IN2P3, Universit\'{e} Paris-Saclay, Orsay}
\affiliation{\'Ecole Polytechnique F\'ed\'erale de Lausanne (EPFL), Lausanne 1015}
\affiliation{P.N. Lebedev Physical Institute of the Russian Academy of Sciences, Moscow 119991}
\affiliation{Liaoning Normal University, Dalian 116029}
\affiliation{Faculty of Mathematics and Physics, University of Ljubljana, 1000 Ljubljana}
\affiliation{Ludwig Maximilians University, 80539 Munich}
\affiliation{Luther College, Decorah, Iowa 52101}
\affiliation{Malaviya National Institute of Technology Jaipur, Jaipur 302017}
\affiliation{University of Malaya, 50603 Kuala Lumpur}
\affiliation{University of Maribor, 2000 Maribor}
\affiliation{Max-Planck-Institut f\"ur Physik, 80805 M\"unchen}
\affiliation{School of Physics, University of Melbourne, Victoria 3010}
\affiliation{University of Mississippi, University, Mississippi 38677}
\affiliation{University of Miyazaki, Miyazaki 889-2192}
\affiliation{Moscow Physical Engineering Institute, Moscow 115409}
\affiliation{Moscow Institute of Physics and Technology, Moscow Region 141700}
\affiliation{Graduate School of Science, Nagoya University, Nagoya 464-8602}
\affiliation{Kobayashi-Maskawa Institute, Nagoya University, Nagoya 464-8602}
\affiliation{Universit\`{a} di Napoli Federico II, 80055 Napoli}
\affiliation{Nara Women's University, Nara 630-8506}
\affiliation{National Central University, Chung-li 32054}
\affiliation{National United University, Miao Li 36003}
\affiliation{Department of Physics, National Taiwan University, Taipei 10617}
\affiliation{H. Niewodniczanski Institute of Nuclear Physics, Krakow 31-342}
\affiliation{Nippon Dental University, Niigata 951-8580}
\affiliation{Niigata University, Niigata 950-2181}
\affiliation{Novosibirsk State University, Novosibirsk 630090}
\affiliation{Osaka City University, Osaka 558-8585}
\affiliation{Pacific Northwest National Laboratory, Richland, Washington 99352}
\affiliation{Panjab University, Chandigarh 160014}
\affiliation{Peking University, Beijing 100871}
\affiliation{Research Center for Nuclear Physics, Osaka University, Osaka 567-0047}
\affiliation{Theoretical Research Division, Nishina Center, RIKEN, Saitama 351-0198}
\affiliation{RIKEN BNL Research Center, Upton, New York 11973}
\affiliation{University of Science and Technology of China, Hefei 230026}
\affiliation{Seoul National University, Seoul 151-742}
\affiliation{Showa Pharmaceutical University, Tokyo 194-8543}
\affiliation{Soongsil University, Seoul 156-743}
\affiliation{University of South Carolina, Columbia, South Carolina 29208}
\affiliation{Stefan Meyer Institute for Subatomic Physics, Vienna 1090}
\affiliation{Sungkyunkwan University, Suwon 440-746}
\affiliation{School of Physics, University of Sydney, New South Wales 2006}
\affiliation{Department of Physics, Faculty of Science, University of Tabuk, Tabuk 71451}
\affiliation{Tata Institute of Fundamental Research, Mumbai 400005}
\affiliation{Department of Physics, Technische Universit\"at M\"unchen, 85748 Garching}
\affiliation{Toho University, Funabashi 274-8510}
\affiliation{Department of Physics, Tohoku University, Sendai 980-8578}
\affiliation{Earthquake Research Institute, University of Tokyo, Tokyo 113-0032}
\affiliation{Department of Physics, University of Tokyo, Tokyo 113-0033}
\affiliation{Tokyo Institute of Technology, Tokyo 152-8550}
\affiliation{Tokyo Metropolitan University, Tokyo 192-0397}
\affiliation{Utkal University, Bhubaneswar 751004}
\affiliation{Virginia Polytechnic Institute and State University, Blacksburg, Virginia 24061}
\affiliation{Wayne State University, Detroit, Michigan 48202}
\affiliation{Yamagata University, Yamagata 990-8560}
\affiliation{Yonsei University, Seoul 120-749}
  \author{R.~Seidl}\affiliation{RIKEN BNL Research Center, Upton, New York 11973} 
  \author{I.~Adachi}\affiliation{High Energy Accelerator Research Organization (KEK), Tsukuba 305-0801}\affiliation{SOKENDAI (The Graduate University for Advanced Studies), Hayama 240-0193} 
  \author{J.~K.~Ahn}\affiliation{Korea University, Seoul 136-713} 
  \author{H.~Aihara}\affiliation{Department of Physics, University of Tokyo, Tokyo 113-0033} 
  \author{D.~M.~Asner}\affiliation{Brookhaven National Laboratory, Upton, New York 11973} 
  \author{V.~Aulchenko}\affiliation{Budker Institute of Nuclear Physics SB RAS, Novosibirsk 630090}\affiliation{Novosibirsk State University, Novosibirsk 630090} 
  \author{T.~Aushev}\affiliation{Moscow Institute of Physics and Technology, Moscow Region 141700} 
  \author{R.~Ayad}\affiliation{Department of Physics, Faculty of Science, University of Tabuk, Tabuk 71451} 
  \author{A.~M.~Bakich}\affiliation{School of Physics, University of Sydney, New South Wales 2006} 
  \author{V.~Bansal}\affiliation{Pacific Northwest National Laboratory, Richland, Washington 99352} 
% ...
\collaboration{The Belle Collaboration}